\documentclass[aps, twocolumn, letterpaper, superscriptaddress, showpacs]{revtex4}

\usepackage{amsmath}
\usepackage{amssymb}
\usepackage{mathrsfs}
\usepackage{xspace}
\usepackage{graphicx}
\usepackage{braket}
\usepackage{color}

\usepackage{ifpdf}
\ifpdf
\fi

\newcommand{\eq}[1]{(\ref{#1})}
\newcommand{\Eq}[1]{Eq.~(\ref{#1})}
\newcommand{\Eqs}[1]{Eqs.~(\ref{#1})}
\newcommand{\Fig}[1]{Fig.~\ref{#1}}
\newcommand{\Sec}[1]{Sec.~\ref{#1}}
\newcommand{\Ref}[1]{Ref.~\cite{#1}}
\newcommand{\Refs}[1]{Refs.~\cite{#1}}
\newcommand{\App}[1]{Appendix~\ref{#1}}

\newcommand{\eg}{{e.g.,\/}\xspace}
\newcommand{\ie}{{i.e.,\/}\xspace}

\newcommand{\mc}[1]{\mathcal{#1}}
\newcommand{\mcc}[1]{\mathfrak{#1}}

\renewcommand{\vec}[1]{{\boldsymbol{\rm #1}}}
\newcommand{\avr}[1]{\left\langle #1 \right\rangle}
\newcommand{\favr}[1]{\langle #1 \rangle}
\newcommand{\pd}{\partial}

\begin{document}
\title{Parametric decay of plasma waves near the upper-hybrid resonance}
\author{I.~Y. Dodin}
\affiliation{Princeton Plasma Physics Laboratory, Princeton, New Jersey 08543, USA}
\author{A.~V. Arefiev}
\affiliation{Center for High Energy Density Science, The University of Texas, Austin, Texas 78712, USA}

\begin{abstract}
An intense X wave propagating perpendicularly to dc magnetic field is unstable with respect to a parametric decay into an electron Bernstein wave and a lower-hybrid wave. A modified theory of this effect is proposed that extends to the high-intensity regime, where the instability rate $\gamma$ ceases to be a linear function of the incident-wave amplitude. An explicit formula for $\gamma$ is derived and expressed in terms of cold-plasma parameters. Theory predictions are in reasonable agreement with the results of the particle-in-cell simulations presented in a separate publication.
\end{abstract}

\pacs{52.35.-g, 52.35.Mw, 52.35.Hr}


\maketitle


\section{Introduction}

The increase in the computer power and advances in numerical modeling have recently made it possible to simulate the propagation of radiofrequency plasma waves using first-principle particle-in-cell (PIC) algorithms. In particular, there has been a growing interest in PIC modeling of the X-B conversion \cite{ref:xiao15, tex:myebw1}, \ie a transformation of an externally-launched electromagnetic X wave into an electron Bernstein wave (EBW), which is useful for depositing energy into the dense core of tokamak plasmas \cite{ref:urban11, ref:fisch87}. In \Ref{tex:myebw1}, an instability was observed in such simulations that results in a parametric decay of an X wave into an EBW and a lower-hybrid wave (LHW). The existing theory of this instability \cite{foot:surkov, ref:gusakov07} is limited to relatively low amplitudes and assumes that the EBW and LHW are inhomogeneous and propagate at nonzero group velocities. (For other relevant studies, see \Refs{ref:mcdermott82, tex:porkolab81, ref:lin81}.) This involves delicate assumptions about how thermal effects enter the wave dispersion relation. But the simulation results in \Ref{tex:myebw1} indicate that the instability is a robust cold-plasma effect and is experienced by homogeneous waves too. Hence, a different theory is needed to explain those results explicitly and also to provide a more robust description of the effect in general. 

Here we propose such theory. By using a variational approach, we derive the instability rate $\gamma$ through the cold-plasma linear susceptibility. This approach is advantageous in the sense that nonlinear ponderomotive forces on the plasma, which are somewhat complicated, do not need to be calculated. Our theoretical predictions for $\gamma$ are in reasonable agreement with simulation results in \Ref{tex:myebw1}. We also extend the theory to higher amplitudes, when the wave interaction is not quite resonant, and $\gamma$ is a nonlinear function of the X-wave amplitude.

The work is organized as follows. In \Sec{sec:notation}, we define the basic notation used in this paper. In \Sec{sec:basic}, we introduce our general approach. In \Sec{sec:simp}, we derive the instability rate. In \Sec{sec:disc}, we present estimates comparing our theory with simulations in \Ref{tex:myebw1} and also summarize the main results of our paper. Some auxiliary calculations are reported in appendices.

\section{Notation}
\label{sec:notation}

The symbol $\doteq$ will be used for definitions. Hats $(\,\hat{\ }\,)$ are used to denote nonlocal (differential or integral) operators. Also, for any $a$, the notation ``$\delta a: $'' will denote that the corresponding equation represents an Euler-Lagrange equation (ELE) obtained by extremizing the action functional $\mc{S}$ with respect to $a$. Also, for any $\smash{E^{(q)}}$, where $q$ denotes the wave type (X, EBW, LHW), we use the following convention for the complex representation: 
\begin{gather}
E^{(q)} = \text{Re}\,E_{\rm c}^{(q)}, \quad E_{\rm c}^{(q)} = \mc{E}^{(q)} e^{i\theta_q}.
\end{gather}	
Here $\smash{\mc{E}^{(q)}}$ is the wave envelope, and $\smash{\theta_q}$ is the rapid phase. The corresponding frequency and wave vector are defined as $\smash{\omega^{(q)} \doteq - \pd_t \theta_q}$ and $\smash{\vec{k}^{(q)} \doteq \nabla \theta_q}$. In particular, when a wave is stationary and homogeneous, one has $\smash{\theta_q (t, \vec{x}) = - \omega^{(q)} t + \vec{k}^{(q)} \cdot \vec{x}} + \text{const}$.

For each given species $s$, $q_s$ and $m_s$ denote the particle charge and mass, $n_s$ is the unperturbed density, $\omega_{ps} \doteq (4\pi n_s q_s^2/m_s)^{1/2}$ is the corresponding plasma frequency, $\Omega_s \doteq q_s B_0/m_s c$ is the cyclotron frequency, $B_0$ is the magnitude of the background dc magnetic field, and $c$ is the speed of light. For simplicity, we consider a single type of ions, denoted with index $_i$, but it is straightforward to generalize the theory to multiple ion types. Electrons are denoted with index $_e$, and we also introduce $e \doteq |q_e|$. The symbols $\omega_{\rm LH}$ and $\omega_{\rm UH}$ denote the lower-hybrid (LH) and upper-hybrid (UH) frequencies (\App{app:cold}). Notably, the constant $\omega_{\rm LH}$ is not necessarily the same as $\omega^{({\rm lh})}$, which is the actual frequency of the LHW that, in general, can depend on the plasma temperature and also on $\smash{\vec{k}^{({\rm lh})}}$.

\section{Basic approach}
\label{sec:basic}

\subsection{Resonance conditions}
\label{sec:rescon}

We consider a process in which an X wave (a ``pump'') scatters into an EBW and a LHW approximately under the conditions of the three-wave resonance: 
\begin{gather}
\omega^{({\rm x})} \approx \omega^{({\rm ebw})} + \omega^{({\rm lh})}, \label{eq:wr}\\
\vec{k}^{({\rm x})} \approx \vec{k}^{({\rm ebw})} + \vec{k}^{({\rm lh})}. \label{eq:kr}
\end{gather}
That being said, we allow the LHW envelope to evolve in time at a rate comparable to $\smash{\omega^{({\rm lh})}}$. The temporal resonance \eq{eq:wr} is assumed satisfied only in the sense that $\smash{\omega^{({\rm x})} \approx \omega^{({\rm ebw})}}$, because, in any case, $\smash{\omega^{({\rm lh})} \sim \omega_{\rm LH} \ll \omega_{\rm UH} \sim \omega^{({\rm ebw})}}$. [The superposition of the X-wave and EBW fields can be considered as a quasimonochromatic field, henceforth called UH field.] In contrast, all the three wave vectors are allowed to be comparable to each other.

Assuming $\smash{\omega^{({\rm lh})}}$ is close to $\omega_{\rm LH}$ irrespective of $\smash{\vec{k}^{({\rm lh})}}$, \Eq{eq:wr} determines $\smash{\omega^{({\rm ebw})}}$. In a one-dimensional (1D) problem, this sets $\smash{\vec{k}^{({\rm ebw})}}$ through the EBW dispersion relation; then, \Eq{eq:kr} sets $\smash{\vec{k}^{({\rm lh})}}$. In this model, \Eqs{eq:wr} and \eq{eq:kr} can always be satisfied exactly. (In multiple dimensions, ensuring the resonance is even easier.) However, when thermal effects are taken into account, the availability of an exact resonance can be a subtle issue, so we allow for nonzero detuning  $\Delta \omega \doteq \smash{\omega^{({\rm x})} - \omega^{({\rm ebw})} - \omega^{({\rm lh})}}$ treated as a free parameter.

\subsection{Variational principle}
\label{sec:vp}

In contrast with \Refs{foot:surkov, ref:gusakov07}, where field equations are derived by calculating ponderomotive on plasma particles, we propose an arguably more transparent formulation in terms of the plasma \textit{linear} susceptibility. The fact that ponderomotive forces can be inferred from the linear susceptibility is widely known, for example, as the $K$-$\chi$ theorem \cite{ref:cary77, ref:kaufman87, my:kchi, tex:myqponder, my:lens}. We adopt a variational approach to utilize this link efficiently. We assume dissipation to be negligible for simplicity, but the general method used here is extendable to dissipative waves too \cite{tex:mynonloc}. It is also to be noted that a related calculation was proposed recently in \Ref{my:shpulse} in application to Raman scattering. (For earlier applications of the variational approach to three-wave interactions, see, \eg \Ref{ref:galloway71}.)

Assuming that dissipation is negligible, the wave interaction can be described using the least action principle $\delta \mc{S} = 0$. The functional $\mc{S}$ can be adopted in the form $\mc{S} = \int \mcc{L}\,dt\,d^3x$ (the Minkowski metric is assumed), and the Lagrangian density $\mcc{L}$ has the form
\begin{gather}
\mcc{L} = \mcc{L}^{({\rm x})} + \mcc{L}^{({\rm ebw})} + \mcc{L}^{({\rm lh})} + \mcc{L}^{({\rm int})},\\
\mcc{L}^{(q)} = \frac{1}{16\pi}\,\vec{E}_{\rm c}^{(q)*} \cdot \hat{\vec{\mcc{D}}} \cdot \vec{E}_{\rm c}^{(q)},\\
\mcc{L}^{({\rm int})} = \frac{1}{16\pi}\,\text{Re}\,\Big\{
\vec{E}_{\rm c}^{({\rm x})*} \cdot \hat{\vec{\chi}}^{({\rm int})}_{\rm c} \cdot \vec{E}_{\rm c}^{({\rm ebw})}\Big\}.
\end{gather}
(If this is not obvious, see \App{app:vp}.) Here, $\smash{\hat{\vec{\mcc{D}}}}$ is the linear dispersion operator given by
\begin{gather}\label{eq:D}
\hat{\vec{\mcc{D}}} \doteq \frac{c^2}{\hat{\omega}^2}\big[
\hat{\vec{k}}\hat{\vec{k}} - \vec{1}(\hat{\vec{k}} \cdot \hat{\vec{k}})\big]
+ \hat{\vec{\epsilon}}_0,
\end{gather}
$\smash{\hat{\vec{\epsilon}}_0}$ is the linear dielectric tensor in the operator form, $\smash{\hat{\vec{\chi}}^{({\rm int})}_{\rm c}}$ is the LHW-driven perturbation to the linear-susceptibility operator, and
\begin{gather}\label{eq:wk}
\hat{\omega} \doteq i \pd_t, \quad \hat{\vec{k}} \doteq -i\nabla.
\end{gather}
Using these, one can consider the dielectric tensor (which is Hermitian in the absence of dissipation \cite{tex:mynonloc}) as a pseudodifferential operator; \ie $\smash{\hat{\vec{\epsilon}}_0 = \vec{\epsilon}_0(t, \vec{x}, \hat{\omega}, \hat{\vec{k}})}$, where $\vec{\epsilon}_0$ is a tensor function. (The prefix ``pseudo'' indicates that the expansion of $\vec{\epsilon}_0$ in $\hat{\omega}$ and $\hat{\vec{k}}$ can contain infinite powers; \ie although expressed in terms of derivatives, such operator can be essentially nonlocal.) Similarly, we introduce a tensor function $\vec{\mcc{D}}$ via $\smash{\hat{\vec{\mcc{D}}} = \vec{\mcc{D}}(t, \vec{x}, \hat{\omega}, \hat{\vec{k}})}$.

Note that ELEs derived from the variational principle are manifestly Lagrangian. In particular, they conserve the total number of high-frequency quanta, as guaranteed by the fact that $\hat{\vec{\mcc{D}}}$ is Hermitian \cite{my:wkin}. It is to be noted that the theory proposed in \Ref{foot:surkov, ref:gusakov07} does not have this property, for it relies on the false assumption that $\smash{\hat{\vec{\epsilon}}_0}$ can be inferred from the homogeneous-plasma dielectric tensor $\smash{\hat{\vec{\epsilon}}_{0,h}}$ simply by replacing $\smash{\vec{k}^{(q)}}$ with $\smash{\hat{\vec{k}}}$. Fixing this issue would require a derivation of $\smash{\hat{\vec{\epsilon}}_0}$ without assuming the geometrical-optics approximation, because $\smash{\hat{\vec{\epsilon}}_{0,h}}$ does not contain enough information in principle \cite{foot:zonal}. It is not our goal to make such revision in the present paper. Instead, we are interested in modifying the theory in other respects (\eg extending it to higher amplitudes), so here we limit our consideration to homogeneous plasmas.

\subsection{Reduced problem}

The simplified system that we study here is as follows. We assume that the background plasma is stationary and homogeneous, so $\mcc{L}$ does not contain explicit dependence on $(t, \vec{x})$. We also consider the X wave as prescribed; then $\smash{\mcc{L}^{({\rm x})}}$ can be dropped. We also adopt the electrostatic approximation for the EBW and LHW (yet not for the X wave), so $\smash{\vec{E}^{({\rm ebw})} \lVert\, \vec{k}^{({\rm ebw})}}$ and $\smash{\vec{E}^{({\rm lh})} \lVert\, \vec{k}^{({\rm lh})}}$; besides, the dispersion operator reduces to $\smash{\hat{\mcc{D}} = \hat{\epsilon}_{0,xx}}$. We also assume 1D propagation along the $x$ axis, which is transverse to the dc magnetic field $\vec{B}_0 = \vec{e}_z B_0$ ($\vec{e}_j$ is a unit vector along the $j$th axis). Then,
\begin{gather}
\mcc{L} = \mcc{L}^{({\rm ebw})} + \mcc{L}^{({\rm lh})} + \mcc{L}^{({\rm int})},\\
\mcc{L}^{(q)} = \frac{1}{16\pi}\,E_{\rm c}^{(q)*}\,\hat{\mcc{D}}  E_{\rm c}^{(q)},\\
\mcc{L}^{({\rm int})} = \frac{1}{16\pi}\,\text{Re}\,\Big\{
\big[\vec{E}_{\rm c}^{({\rm x})*} \cdot \hat{\vec{\chi}}^{({\rm int})}_{\rm c} \cdot \vec{e}_x\big]E_{\rm c}^{({\rm ebw})}\Big\}.
\end{gather}

Since $\mcc{L}^{({\rm int})}$ is small, it is enough to calculate $\smash{\hat{\vec{\chi}}^{({\rm int})}_{\rm c}}$ approximately, so we adopt the cold-plasma approximation for that. In this approximation, the susceptibility is entirely determined by the particle densities and by the magnetic field. But the LHW is assumed electrostatic, so it does not perturb the magnetic field. Thus,
\begin{gather}\label{eq:ch}
\hat{\vec{\chi}}^{({\rm int})}_{\rm c}
 = \sum_s \frac{\pd \hat{\vec{\chi}}_s^{({\rm uh})}}{\pd n_s}\,n^{({\rm lh})}_{s,c}
 = \sum_s \frac{n^{({\rm lh})}_{s,c}}{n_s}\,\hat{\vec{\chi}}_s^{({\rm uh})},
\end{gather}
where the summation is taken over species, and $\smash{\hat{\vec{\chi}}_s^{({\rm uh})}}$ are the corresponding unperturbed susceptibility operators acting on the UH field; hence the index $\smash{^{({\rm uh})}}$. Assuming the UH field is quasimonochromatic, we can approximate them with $\smash{\vec{\chi}_s(\omega_{\rm UH})}$. Also, $n_s$ are the corresponding unperturbed susceptibility operators and unperturbed densities, and $\smash{n^{({\rm lh})}_{s,c}}$ are the LH density perturbations in the complex representation. From the corresponding continuity equations, the latter are found to be
\begin{gather}
n^{({\rm lh})}_{s,{\rm c}} =  \frac{\hat{k}^{({\rm lh})} \hat{\chi}_{s,xx}^{({\rm lh})}}{4\pi i q_s}\,E_{\rm c}^{({\rm lh})}.
\end{gather}
We added the index $\smash{^{({\rm lh})}}$ to emphasize that the corresponding operators act on the LH field. Since $\smash{\hat{\chi}^{({\rm lh})}_{e,xx} \sim \hat{\chi}^{({\rm lh})}_{i,xx}}$ (\App{app:cold}), one has $n_e \sim n_i$. But $\smash{\hat{\vec{\chi}}_e^{({\rm uh})} \gg \hat{\vec{\chi}}_i^{({\rm uh})}}$. This means that, in \Eq{eq:ch}, it is enough to retain just the electron contribution. (In other words, the ponderomotive force on ions is neglected.) That gives
\begin{gather}\label{eq:ch2}
\hat{\vec{\chi}}^{({\rm int})}_{\rm c} \approx \frac{\hat{k}^{({\rm lh})} \hat{\chi}_{e,xx}^{({\rm lh})}}{4\pi i q_e n_e}\,E_{\rm c}^{({\rm lh})}\,\vec{\chi}_s(\omega_{\rm UH}).
\end{gather}
Hence,
\begin{gather}
\mcc{L}^{({\rm int})} \approx \frac{1}{16\pi}\,\text{Re}\,\Big\{2i\mc{E}^{({\rm ebw})}
\big[e^{i\theta_{\text{ebw}} - i \theta_{\text{x}}}\hat{\beta}E_{\rm c}^{({\rm lh})}\big]
\Big\},\\
\hat{\beta} \doteq -\big[\vec{\mc{E}}^{({\rm x})*} \cdot \vec{\chi}_s(\omega_{\rm UH}) \cdot \vec{e}_x\big]\,\frac{\hat{k}^{({\rm lh})} \hat{\chi}_{e,xx}^{({\rm lh})}}{8\pi q_e n_e}.
\end{gather}
Using \Eq{eq:brck}, one can simplify the expression in the square brackets down to $\smash{-\mc{E}_x^{({\rm x})*}}$. Also, $\smash{\hat{\chi}_{e,xx}^{({\rm lh})} \approx \omega_{pe}^2/\Omega_e^2}$, which is just a constant. Below, we also assume for simplicity that waves are spatially monochromatic. This implies that the condition of the spatial resonance \eq{eq:kr} is satisfied exactly and $\smash{\hat{k}E_{\rm c}^{(q)} = k^{(q)} E_{\rm c}^{(q)}}$. Hence, the operator $\smash{\hat{\beta}}$ can be replaced with the following constant:
\begin{gather}
\beta = \frac{k^{({\rm lh})}}{8\pi q_e n_e}\,\frac{\omega_{pe}^2}{\Omega_e^2}.
\end{gather}
Then, the formula for $\mcc{L}^{({\rm int})}$ is summarized as follows:
\begin{gather}
\mcc{L}^{({\rm int})} \approx \frac{1}{16\pi}\,\text{Re}\,\Big[2i\beta\mc{E}^{({\rm ebw})} \mc{E}^{({\rm lh})} e^{-i(\theta_{\text{x}} - \theta_{\text{ebw}} - \theta_{\text{lh}})}\Big].
\end{gather}

\section{Instability rate}
\label{sec:simp}

\subsection{Weak pump}
\label{sec:weak}

First, suppose that the pump amplitude $\smash{\mc{E}^{({\rm x})}}$ is small enough. Then, the EBW and the LHW oscillate approximately at the unperturbed frequencies that satisfy
\begin{gather}\label{eq:ldr}
\mcc{D}(\omega^{(q)}, k^{(q)}) = 0.
\end{gather}
Accordingly, we can use
\begin{gather}\notag
e^{-i\theta_q} \hat{\mcc{D}} E_{\rm c}^{(q)} =
\mcc{D}\big(\omega^{(q)} + \hat{\omega}, k^{(q)}\big) \mc{E}^{(q)}
\approx \mcc{D}_\omega^{(q)} \hat{\omega} \mc{E}^{(q)},
\end{gather}
where $\mcc{D}_\omega^{(q)} \doteq \pd_\omega \mcc{D}(\omega^{(q)}, k^{(q)})$. This gives
\begin{multline}\notag
16\pi \mcc{L} = \mc{E}^{({\rm ebw})*} \mcc{D}_\omega^{({\rm ebw})} i\pd_t \mc{E}^{({\rm ebw})} + \mc{E}^{({\rm lh})*} \mcc{D}_\omega^{({\rm lh})} i\pd_t \mc{E}^{({\rm lh})} \\
+ i\beta \mc{E}^{({\rm ebw})} \mc{E}^{({\rm lh})} e^{i \Delta \omega t} - i\beta^* \mc{E}^{({\rm ebw})*} \mc{E}^{({\rm lh})*} e^{-i \Delta \omega t}.
\end{multline}
The corresponding ELEs are as follows:
\begin{align}
\delta \mc{E}^{({\rm ebw})}: \quad & 
\pd_t \mc{E}^{({\rm ebw})*} = \frac{\beta \mc{E}^{({\rm lh})}}{\mcc{D}_\omega^{({\rm ebw})}}\,
e^{i \Delta \omega t},
\\
\delta \mc{E}^{({\rm lh})*}: \quad & \pd_t \mc{E}^{({\rm lh})}
= \frac{\beta^*\mc{E}^{({\rm ebw})*}}{\mcc{D}_\omega^{({\rm lh})}}\,
e^{-i \Delta \omega t}
\end{align}
(plus the two equations adjoint to these, which we do not need to consider). These can be combined into a single equation for $\mc{E}^{({\rm lh})}$:
\begin{gather}
\pd^2_t \mc{E}^{({\rm lh})} + i (\Delta \omega) \pd_t \mc{E}^{({\rm lh})} - \gamma_0^2 \mc{E}^{({\rm lh})} = 0,
\end{gather}
where $\gamma_0$ is a constant given by
\begin{gather}\label{eq:gamma0}
\gamma_0 \doteq \frac{|\beta|}{\sqrt{\mcc{D}_\omega^{({\rm ebw})}\mcc{D}_\omega^{({\rm lh})}}}.
\end{gather}
In particular, for $\mc{E}^{({\rm lh})} \propto \exp(-i\omega t)$, one gets
\begin{gather}\label{eq:w12}
\omega_{1,2} = \frac{\Delta \omega}{2} \pm \sqrt{
\frac{(\Delta \omega)^2}{4} - \gamma_0^2}.
\end{gather}
Clearly, $\gamma_0$ is real, because the sign of $\mcc{D}_\omega^{(q)}$ coincides with that of the mode energy \cite{my:amc} and the waves involved have positive energies. Hence, if $\Delta \omega < 2\gamma_0$, an instability develops with the rate
\begin{gather}\label{eq:gamma}
\gamma = \sqrt{\gamma_0^2 - \frac{(\Delta \omega)^2}{4}}.
\end{gather}
In order to explicitly calculate $\gamma_0$ that enters here, we invoke \Eqs{eq:ldr}, \eq{eq:dB}, and \eq{eq:dLH}. This leads to
\begin{gather}
\mcc{D}^{({\rm ebw})}_\omega \approx \frac{2\omega_{\rm UH}}{\omega_{pe}^2},
\quad
\mcc{D}^{({\rm lh})}_\omega \approx \frac{2\omega_{pi}^2}{\omega_{\rm LH}^3},
\end{gather}
where we neglected thermal corrections. We also invoke the condition of plasma neutrality, namely, $\smash{\omega_{pe}^2/\Omega_e} = - \smash{\omega_{pi}^2/\Omega_i}$. Then, \Eq{eq:gamma0} gives
\begin{gather}\label{eq:gamma0ans}
\gamma_0 \approx \omega_{\rm LH}\,\frac{\omega_{pe}\omega_{pi}}{|\Omega_e\Omega_i|}
\sqrt{\frac{\omega_{\rm LH}}{\omega_{\rm UH}}}\,\frac{\big|k^{({\rm lh})} \mc{E}^{({\rm x})}_x\big|}{16 \pi e n_e}.
\end{gather}

Although the effect calculated here is due to the existence of EBW, which implies a nonzero temperature $T$, the instability rate \eq{eq:gamma} is insensitive to $T$ (except, of course, at large enough $T$). In this sense, the instability can be understood as a cold-plasma effect. Also note that, since $\smash{\gamma_0 \propto |k^{({\rm lh})}| = |k^{({\rm x})}_x - k^{({\rm ebw})}_x|}$, the instability rate is somewhat larger for backscattering, when $\smash{k^{({\rm x})}_x}$ and $\smash{k^{({\rm ebw})}_x}$ have opposite signs.

\subsection{Strong pump}

Now let us consider the case when the pump is strong enough so the LHW ceases to be quasimonochromatic. (The EBW is still assumed quasimonochromatic because it has a much higher frequency.) In this case, we adopt the Lagrangian density in the form
\begin{multline}\notag
16\pi \mcc{L} = 
\mc{E}^{({\rm ebw})*} \mcc{D}_\omega^{({\rm ebw})} i\pd_t \mc{E}^{({\rm ebw})} + \bar{\mc{E}}^{({\rm lh})*} \hat{\mcc{D}} \bar{\mc{E}}^{({\rm lh})} \\
+ i\beta \mc{E}^{({\rm ebw})} \bar{\mc{E}}^{({\rm lh})} e^{i \vartheta t} - i\beta^* \mc{E}^{({\rm ebw})*} \bar{\mc{E}}^{({\rm lh})} e^{-i \vartheta t}.
\end{multline}
Here, $\smash{\vartheta \doteq \omega^{({\rm x})} - \omega^{({\rm ebw})} = \omega^{({\rm lh})} + \Delta \omega}$, and we introduced $\smash{\bar{\mc{E}}^{({\rm lh})}(t) \doteq E_c^{({\rm lh})}(t, x) \exp(-i \vec{k}^{(\rm lh)} \cdot \vec{x})}$ that is not necessarily slow in time but has no dependence on $x$. Rather than expanding the dispersion operator for the LHW, as in \Sec{sec:weak}, we now adopt the fully nonlocal operator inferred from \Eq{eq:dLH}:
\begin{gather}
\hat{\mcc{D}}
=\hat{\omega}^{-2}(\hat{\omega}^2 - \omega_{\rm LH}^2)\,
\omega_{pi}^2/\omega_{\rm LH}^2.
\end{gather}

The corresponding ELEs are as follows:
\begin{align}
\delta \mc{E}^{({\rm ebw})}: \quad & 
\pd_t \mc{E}^{({\rm ebw})*} = \frac{\beta \bar{\mc{E}}^{({\rm lh})}}{\mcc{D}_\omega^{({\rm ebw})}}\,
e^{i\vartheta t},
\label{eq:aux3}\\
\delta \bar{\mc{E}}^{({\rm lh})*}: \quad & \hat{\mcc{D}}\bar{\mc{E}}^{({\rm lh})}
= i\beta^*\mc{E}^{({\rm ebw})*}\,
e^{-i\vartheta t}\label{eq:aux4}
\end{align}
(plus the two equations adjoint to these, which we do not need to consider). Using \Eq{eq:aux3}, we get
\begin{gather}
 \mc{E}^{({\rm ebw})*} e^{-i\vartheta t} = \frac{i \beta}{\mcc{D}_\omega^{({\rm ebw})}}\, (\hat{\omega} - \vartheta)^{-1} \bar{\mc{E}}^{({\rm lh})}.
\end{gather}
Accordingly, \Eq{eq:aux4} can be expressed as
\begin{gather}
\hat{\mcc{D}}\bar{\mc{E}}^{({\rm lh})}
= -\frac{|\beta|^2}{\mcc{D}_\omega^{({\rm ebw})}}\,(\hat{\omega} - \vartheta)^{-1} \bar{\mc{E}}^{({\rm lh})},
\end{gather}
or, equivalently,
\begin{gather}
\big[(\hat{\omega} - \vartheta)(\hat{\omega}^2 - \omega_{\rm LH}^2) + 2\gamma_0^2 \hat{\omega}^2\big]\bar{\mc{E}}^{({\rm lh})} = 0,
\end{gather}
where $\gamma_0$ is given by \Eq{eq:gamma0ans}. The corresponding dispersion relation is as follows:
\begin{gather}\label{eq:cubic}
(\omega - \vartheta)(\omega^2 - \omega_{\rm LH}^2) + 2\gamma_0^2 \omega^2 = 0.
\end{gather}

\begin{figure}
\centering
\includegraphics[width=.49\textwidth]{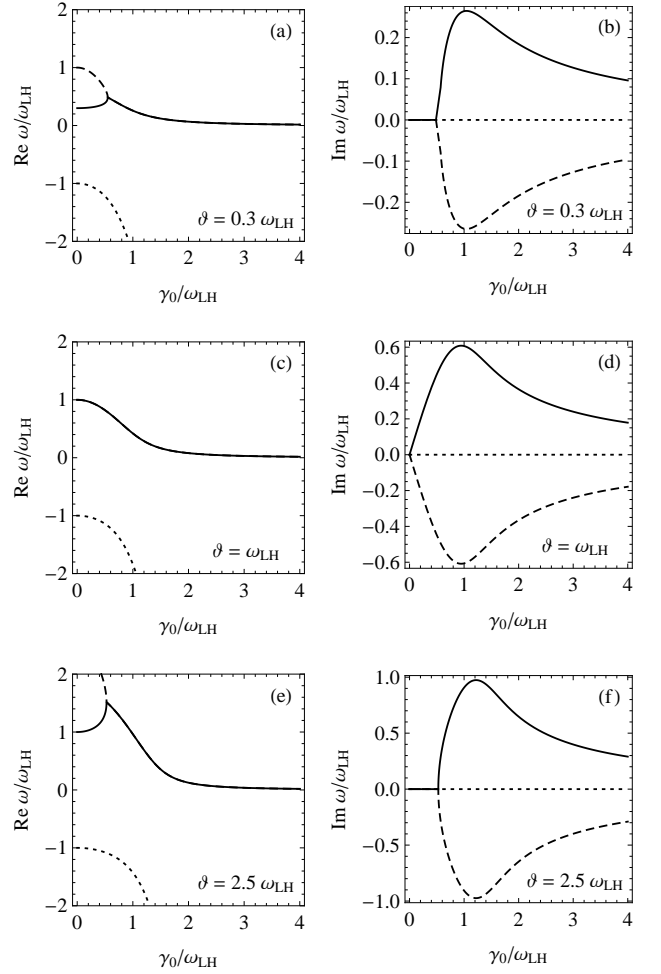}
\caption{Solutions of \Eq{eq:cubic} for $\omega$ versus $\gamma_0$, both measured in units $\omega_{\rm LH}$. The left column shows the real part of $\omega$, and the right column shows the imaginary part of $\omega$. In (a) and (b), $\vartheta = 0.3\omega_{\rm LH}$. In (c) and (d), $\vartheta = \omega_{\rm LH}$. In (e) and (f), $\vartheta =2.5\omega_{\rm LH}$. In each given plot, different curves show different branches. The unstable branch is shown as a solid line.}
\label{fig:domain}
\end{figure}  

Equation \eq{eq:cubic}, in combination with \Eq{eq:gamma0ans} for $\gamma_0$, is our main result. It is a cubic equation for $\omega$ and has an exact, albeit cumbersome, analytic solution. We do not present it here for brevity, but see numerical plots in \Fig{fig:domain}. Like in the case of resonant damping, taking $\Delta \omega = 0$ (which corresponds to $\vartheta = \omega_{\rm LH}$) ensures that there is no instability threshold within the adopted model. In other words, even an arbitrarily small $\gamma_0$ causes one of the three roots of \Eq{eq:cubic} to acquire a positive imaginary part. 

One can also derive asymptotic expressions for $\omega$ at $\Delta \omega = 0$ in terms of the dimensionless parameter
\begin{gather}\label{eq:Gamma}
g \doteq \frac{\gamma_0}{\omega_{\rm LH}} \approx \frac{\omega_{pe}\omega_{pi}}{|\Omega_e\Omega_i|}
\sqrt{\frac{\omega_{\rm LH}}{\omega_{\rm UH}}}\,\frac{\big|k^{({\rm lh})} \mc{E}^{({\rm x})}_x\big|}{16 \pi e n_e}.
\end{gather}
Specifically, for $g \gg 1$ (strong pump) one gets
\begin{gather}\label{eq:largeG}
\omega_{1,2}/\omega_{\rm LH} \approx \pm i/(g\sqrt{2}), \quad \omega_3/\omega_{\rm LH} \approx - 2g^2 + 1,
\end{gather}
and for $g \ll 1$ (weak pump) one gets
\begin{gather}\label{eq:smallG}
\omega_{1,2}/\omega_{\rm LH} \approx 1 \pm i g, \quad \omega_3/\omega_{\rm LH} \approx -1 - g^2/2.
\end{gather}
The first two roots in \Eq{eq:smallG} correspond to the two roots predicted by \Eq{eq:w12}. [The additional real part of the frequency that is predicted by \Eq{eq:smallG}, namely, $\omega_{\rm LH}$, is due to the fact that here we derived the frequency of $\smash{\bar{\mc{E}}^{({\rm lh})}}$ rather than that of $\smash{\mc{E}^{({\rm lh})}}$.] Thus, the weak-pump model discussed in \Sec{sec:weak} is successfully recovered as an asymptotic limit of the general theory presented here.

It is to be noted that our general theory differs from its low-amplitude limit in that it takes an additional branch of the dispersion relation into account. Although this branch remains stable by itself [see $\omega_3$ in \Eqs{eq:largeG} and \eq{eq:smallG}], it still affects the stability of resonant branches. This can be identified as a manifestation of polarization (``spin'') effects that were recently discussed for both classical and quantum waves in \Refs{tex:mycovar, my:qdirac}.  

\section{Discussion}
\label{sec:disc}

For an estimate, let us adopt the parameters used in \Ref{tex:myebw1}. Namely, consider an electron-deuterium plasma with electron density $n_e = 0.89 \times 10^{18}\,\text{m}^{-1}$ and temperature $T = 950\,\text{eV}$. Also, $B_0 = 0.25\,\text{T}$, $k^{({\rm lh})} = 2000\,\text{m}^{-1}$, and $\mc{E}^{({\rm x})}_x = 8\times 10^5\,\text{V}/\text{m}$. The X wave is launched at $\omega = 2\pi \times 10\,\text{GHz}$, which is about $0.9\omega_{\rm UH}$, and $\omega_{\rm UH} = 6.9\times 10^{10}\,\text{s}^{-1}$. Also, $\omega_{\rm LH} = 7.9 \times 10^{8}\,\text{s}^{-1}$, which is about $33.1\Omega_i$. The corresponding LH period is $2\pi/\omega_{\rm LH} \approx 7.9\,\text{ns}$. Considering that we neglected thermal effects, which somewhat modify the plasma dispersion, this is in reasonable agreement with the simulation results in \Ref{tex:myebw1}, where oscillations were observed with period about $10\,\text{ns}$. For the specified parameters, \Eq{eq:gamma0ans} gives $\gamma_0 \approx 1.3\times 10^8\,\text{s}^{-1}$. This corresponds to $\gamma_0 \approx 5.5\Omega_i$, or $g \approx 0.17$. Then, the instability rate is expected to be $\gamma \approx \gamma_0$, which corresponds to $\gamma^{-1} \approx 7.6\,\text{ns}$. This result is in the ballpark of the simulation results.

In summary, we proposed a modified theory of the instability that is caused by the resonant scattering, or parametric decay, of an intense X wave into an EBW and LHW. Our theory extends to the high-intensity regime, where the instability rate $\gamma$ ceases to be a linear function of the incident-wave amplitude. We derived an explicit formula for $\gamma$ and expressed it in terms of cold-plasma parameters. Predictions of our theory are in reasonable agreement with the results of the PIC simulations presented in \Ref{tex:myebw1}.

The work was supported by the U.S. DOE through Contract No. DE-AC02-09CH11466 and by the U.S. DOE-NNSA Cooperative Agreement No. DE-NA0002008.

\appendix
\section{Basic properties of the relevant plasma waves}
\label{app:cold}

Here, we summarize some basic properties of the plasma waves relevant to the discussion in the main text.

\subsection{Basic equations}

We consider monochromatic waves in the model of cold stationary homogeneous plasma. The dc magnetic field is adopted in the form $\vec{B}_0 =\vec{e}_z B_0$. Then, the plasma dielectric tensor can be expressed as follows \cite{book:stix}:
\begin{gather}
\vec{\epsilon} = \left(
\begin{array}{ccc}
S & - iD & 0\\
iD & S & 0 \\
0 & 0 & P
\end{array}
\right),
\end{gather}
where $S$, $D$, and $P$ depend on the wave frequency $\omega$ but not on the wave vector. Specifically,
\begin{gather}\notag
S = 1 + \sum_s \chi_{s,xx}, 
\quad
iD = \sum_s \chi_{s,yx},
\quad
P = 1 + \sum_s \chi_{s, zz},
\end{gather}
and $\chi_{s,xy} = \chi_{s,yx}^* = - \chi_{s,yx}$, where
\begin{gather}
\chi_{s, xx} = - \frac{\omega_{ps}^2}{\omega^2 - \Omega_s^2}, \\
\chi_{s, yx} = \frac{i\Omega_s}{\omega}\,\frac{\omega_{ps}^2}{\omega^2 - \Omega_s^2},\\
\chi_{s, zz} = - \frac{\omega_{ps}^2}{\omega^2}.
\end{gather}

For waves propagating perpendicularly to $\vec{B}_0$, the linear field equation $\smash{\hat{\vec{\mcc{D}}}\cdot \vec{E}_c = 0}$ becomes
\begin{gather}\label{eq:fe}
\left(
\begin{array}{ccc}
S & - iD & 0\\
iD & S - N^2 & 0 \\
0 & 0 & P
\end{array}
\right)
\left(
\begin{array}{c}
\mc{E}_x\\
\mc{E}_y\\
\mc{E}_z
\end{array}
\right) = 0,
\end{gather}
where $N \doteq ck/\omega$ is the refraction index. Thus,
\begin{gather}\label{eq:uhpol}
S \mc{E}_x = iD \mc{E}_y.
\end{gather}
Also, the extraordinary waves that are of interest in this paper are defined as those with $\mc{E}_z = 0$, so \Eq{eq:fe} is simplified down to
\begin{gather}
\left(
\begin{array}{cc}
S & - iD\\
iD & S - N^2
\end{array}
\right)
\left(
\begin{array}{c}
\mc{E}_x\\
\mc{E}_y
\end{array}
\right) = 0.
\end{gather}
The corresponding dispersion relation is $N^2 = (S^2-D^2)/S$. The electrostatic limit corresponds to $S = 0$, which equation defines the hybrid resonances.

\subsection{UH waves}

In the UH range, the ion contribution to the dielectric tensor is negligible, so
\begin{gather}
\chi_{xx} \approx \chi_{e,xx} = - \frac{\omega_{pe}^2}{\omega^2 - \Omega_e^2}, \\
\chi_{yx} \approx \chi_{e,yx} = \frac{i\Omega_e}{\omega}\,\frac{\omega_{pe}^2}{\omega^2 - \Omega_e^2},\\
\chi_{zz} \approx \chi_{e,zz} = - \frac{\omega_{pe}^2}{\omega^2}.
\end{gather}
In particular, this implies
\begin{gather}
\vec{\chi}_e \approx \left(
\begin{array}{ccc}
S-1 & - iD & 0\\
iD & S-1 & 0 \\
0 & 0 & P-1
\end{array}
\right),
\end{gather}
so, as seen easily, \Eq{eq:uhpol} leads to
\begin{gather}\label{eq:brck}
\vec{\mc{E}}^{({\rm x})*} \cdot \vec{\chi}_e(\omega_{\rm UH}) \cdot \vec{e}_x = -\mc{E}_x^{({\rm x})*}.
\end{gather}

Since $\smash{S \approx 1 - \omega_{pe}^2/(\omega^2 - \Omega_e^2)}$, the UH frequency is $\smash{\omega_{\rm UH} = (\omega_{pe}^2 + \Omega_e^2)^{1/2}}$. This determines the frequency range for electrostatic EBW, although calculating the EBW dispersion relation requires taking thermal effects into account. In particular, the dispersion relation of the EBW in the electrostatic approximation is $\smash{\mcc{D}^{({\rm ebw})}(\omega, k) = 0}$, where \cite{book:stix}
\begin{gather}
\mcc{D}^{({\rm ebw})}(\omega, k) = 1 - 
\sum_{n=1}^\infty \frac{n^2 \omega_{pe}^2}{\omega^2 - (n\Omega_e)^2}\,\frac{2I_n(\lambda_e)}{\lambda_e}\,e^{-\lambda_e},
\end{gather}
$I_n$ the modified Bessel function of order $n$, $\lambda_e \doteq (k v_{Te}/\Omega_e)^2$, and $v_{Te}$ is the electron thermal speed. For simplicity, we consider the interaction with the lowest-order mode only and adopt
\begin{gather}\label{eq:dB}
\mcc{D}^{({\rm ebw})}(\omega, k) \approx 1 - \frac{\omega_{pe}^2}{\omega^2 - \Omega_e^2}\,\Theta(\lambda_e),\\
\Theta(\lambda_e) \doteq \frac{2I_n(\lambda_e)}{\lambda_e}\,e^{-\lambda_e} = 1 - \lambda_e + \mc{O}(\lambda_e^2),
\end{gather}
using that $\lambda_e$ is small. Note that $\omega^{({\rm ebw})}(\lambda_e \to 0) \to \omega_{\rm UH}$. For parameters listed in \Sec{sec:disc}, one has $\lambda_e \approx 0.35 \lesssim 1$, so neglecting thermal effects is a reasonable approximation.

\subsection{LH waves}

In the LH range ($\Omega_i \ll \omega \ll \Omega_e$), one has
\begin{align}
\chi_{i,xx} \approx - \frac{\omega_{pi}^2}{\omega^2}, &
\quad
\chi_{e,xx} \approx \frac{\omega_{pe}^2}{\Omega_e^2}, \\
\chi_{i,yx} \approx \frac{i\Omega_i}{\omega}\,\frac{\omega_{pi}^2}{\omega^2}, &
\quad
\chi_{e,yx} \approx - \frac{i\Omega_e}{\omega}\,\frac{\omega_{pe}^2}{\Omega_e^2}.
\end{align}
Accordingly, $S \approx 1 + \omega_{pe}^2/\Omega_e^2 - \omega_{pi}^2/\omega^2$, and
\begin{gather}
\omega_{\rm LH} = \frac{\omega_{pi}}{\sqrt{1 + \omega_{pe}^2/\Omega_e^2}}.
\end{gather}
Assuming $\omega_{pe} \sim \Omega_e$, this gives $\omega_{\rm LH} \sim \omega_{pi} \sim \Omega_e (\omega_{pi}/\omega_{pe}) \sim |\Omega_i\Omega_e|^{1/2}$. Hence, at the LHW frequency,
\begin{gather}
\frac{\chi_{i,xx}}{\chi_{e,xx}} \sim 1,
\quad
\frac{\chi_{i,yx}}{\chi_{e,yx}} \ll 1,
\quad 
\frac{\chi_{e,xx}}{\chi_{e,yx}} \ll 1,
\end{gather}
because
\begin{gather}
\frac{\chi_{i,yx}}{\chi_{e,yx}} \sim \frac{\Omega_i}{\Omega_e}\,
\frac{\omega_{pi}^2}{\omega_{pe}^2}\,\frac{\Omega_e^2}{\omega_{\rm LH}^2}
\sim \frac{m_e}{m_i},\\
\frac{\chi_{e,xx}}{\chi_{e,yx}} \sim 
\frac{\omega_{pi}^2}{\Omega_e^2}\, \frac{\omega_{\rm LH}}{\Omega_e}\,
\frac{\Omega_e^2}{\omega_{pe}^2}
\sim \left(
\frac{m_e}{m_i}
\right)^{3/2}.
\end{gather}

Like with EBWs, calculating the dispersion function of electrostatic LHW (ion Bernstein waves) requires that thermal corrections be taken into account. This leads to
\begin{gather}\label{eq:Di}
\mcc{D}^{({\rm lh})}(\omega, k) \approx 1 + \frac{\omega_{pe}^2}{\Omega_e^2} - \sum_{n=1}^\infty \frac{n^2 \omega_{pi}^2}{\omega^2 - (n\Omega_i)^2}\,\frac{2I_n(\lambda_i)}{\lambda_i}\,e^{-\lambda_i},
\end{gather}
where $\lambda_i \doteq (k v_{Ti}/\Omega_i)^2$, and $v_{Ti}$ is the ion thermal speed. In the regime of interest, $\lambda_i \gg 1$. Nevertheless, the sum in \Eq{eq:Di} can be approximated with its cold limit, except when $\omega$ is particularly close to one of cyclotron resonances (\App{app:aux}). Hence, we adopt
\begin{align}
\mcc{D}^{({\rm lh})}(\omega, k) 
& \approx 1 + \frac{\omega_{pe}^2}{\Omega_e^2} - \frac{\omega_{pi}^2}{\omega^2} \notag\\
& = \bigg(1 + \frac{\omega_{pe}^2}{\Omega_e^2}\bigg)\bigg(1 - \frac{\omega_{\rm LH}^2}{\omega^2}\bigg)\notag\\
& = \frac{\omega_{pi}^2}{\omega_{\rm LH}^2}\bigg(1 - \frac{\omega_{\rm LH}^2}{\omega^2}\bigg).
\label{eq:dLH}
\end{align}

\section{Applicability of the cold approximation for $\boldsymbol{\chi_{xx}}$}
\label{app:aux}

The numerical parameters adopted in this paper (\Sec{sec:disc}) correspond to $\smash{\lambda_i \doteq (k v_{Ti}/\Omega_i)^2 \approx 317}$. This number is far too large to allow for an asymptotic small-argument expansion of the modified Bessel functions in \Eq{eq:Di}. Likewise, the large-argument expansion is inapplicable because of the large value of $a \doteq \omega/\Omega_i \approx 33.1$, which determines the number of relevant harmonics ($n \sim a$). Thus, a different approach is needed to justify the cold-plasma approximation \eq{eq:dLH}. It is the purpose of this appendix to provide such justification.

We start by adopting an alternative expression for $\chi_{xx}$ that is equivalent to that assumed in \Eq{eq:Di} but does not involve an infinite series \cite{ref:qin07b}:
\begin{gather}\label{eq:chiT}
\chi_{xx} = \frac{\omega_p^2}{\omega \Omega} \int v_\perp\,
\frac{\pd f_0(v)}{\pd v_\perp}\,
\mcc{T}_{xx}\,2\pi v_\perp\,dv_\perp\,dv_\lVert,\\
\mcc{T}_{xx} = \frac{a}{z^2}\left[
\frac{\pi a}{\sin(\pi a)}\,J_{-a}(z)J_a(z) - 1
\right].\label{eq:Txx}
\end{gather}
(The species index is henceforth dropped for brevity.) Here, $f_0$ is the unperturbed distribution that is assumed isotropic and is normalized such that $\int f_0(v)\, 2\pi v_\perp\,dv_\perp\,dv_\lVert = 1$, $v_\perp$ is the velocity perpendicular to the magnetic field, $v_\lVert$ is the velocity parallel to the magnetic field, and $\smash{v \doteq (v_\perp^2 + v_\lVert^2)^{1/2}}$. Also, $J_a$ are Bessel functions of order $a$. 

Suppose $z$ is small in a sense that is yet to be defined. Then, $\mcc{T}_{xx}$ has the following asymptotic expansion \cite{foot:math}:
\begin{align}
\mcc{T}_{xx} 
& = \sum_{m=0}^\infty
\frac{(-1)^{m+1} a^2 \sqrt{\pi}\, \Gamma(3/2 + m) z^{2m}}
     {\Gamma(2+m) \Gamma(2-a+m) \Gamma (2+a+m) \sin (\pi  a)} \notag\\
& = \sum_{m=0}^\infty  
\frac{m+1/2}{m+1}\frac{(2m)!}{4^m(m!)^2}\,\frac{a z^{2m}}{\left[a^2-(m+1)^2\right]\ldots\left(a^2-1\right)}, \notag
\end{align}
where $\Gamma$ is the gamma function. More explicitly,
\begin{multline}
\mcc{T}_{xx} = 
\frac{a}{2 \left(a^2-1\right)}+\frac{3 a z^2}{8 (a^2-1)(a^2-2^2)}\\
+\frac{5 a z^4}{16(a^2-1)(a^2-2^2)(a^2-3^2)} + ... \label{eq:Taxx}
\end{multline}
In our case, $a \gg 1$, so the ratio of the neighboring terms scales as $\smash{z^2/a^2}$, provided that $a$ is not too close to an integer. Hence, the expansion is applicable roughly at $\smash{z^2 \ll a^2}$. [This is less restrictive than the validity requirement for the small-argument expansion of $J_a(z)$, which is $\smash{z^2 \ll a}$.] By substituting \Eq{eq:Taxx} into \Eq{eq:chiT} and assuming the Maxwellian distribution, we~get
\begin{align}
\chi_{xx} 
& =- \frac{\omega_p^2}{\Omega^2} \sum_{m=0}^\infty
\frac{(-2)^{m+1} a \sqrt{\pi}\, \Gamma(3/2 + m)\lambda^m}
     {\Gamma(2-a+m) \Gamma(2+a+m) \sin(\pi a)} \notag\\
& = - \frac{\omega_p^2}{\Omega^2} \sum_{m=0}^\infty\frac{(2m+1)!}{2^m m!}
     \frac{\lambda^m}{[a^2-(m+1)^2]\ldots(a^2-1)},\notag
\end{align}
or, more explicitly,
\begin{multline}
\chi_{xx} 
=- \frac{\omega_p^2}{\Omega^2}\bigg[
\frac{1}{a^2 - 1}+\frac{3 \lambda}{(a^2 - 1)(a^2 - 2^2)}\\
+\frac{15 \lambda ^2}{(a^2-1)(a^2-2^2)(a^2-3^2)} + ...
\bigg].\label{eq:chimain}
\end{multline}
This expansion is somewhat known \cite{ref:sitenko57}, but here we emphasize not the expansion \textit{per~se} but rather its validity domain. The ratio of the neighboring terms away from resonances scales as $\lambda/a^2$, so the expansion requires, roughly, $\lambda/a^2 \ll 1$. (This is less restrictive compared to the usual requirement $\lambda \ll 1$.) In particular, the cold-plasma limit
\begin{gather}
\chi_{xx} \approx 
- \frac{\omega_p^2}{\Omega^2}
\frac{1}{a^2 - 1} = - \frac{\omega_p^2}{\omega^2 -\Omega^2}
\end{gather}
is reproduced when the second term in the expansion \eq{eq:chimain} is negligible compared to the first term. This implies $3\lambda/a^2 \ll 1$. For parameters adopted in this paper, $3\lambda/a^2 \approx 0.87$, so one can expect the cold-plasma approximation to be applicable at least semiquantitatively.

\section{Variational principle}
\label{app:vp}

In this appendix, we derive the Lagrangian density $\mcc{L}$ introduced in \Sec{sec:vp}. This $\mcc{L}$ consists of the Lagrangian density of the electromagnetic field $\mcc{L}_{\rm em} = (E^2 - B^2)/(8\pi)$ (here and further, $\vec{E}$ denotes the wave electric field, and $\vec{B}$ denotes the wave magnetic field) plus the Lagrangian density of the plasma interaction with this field, $\mcc{L}_p$. Parts of $\mcc{L} = \mcc{L}_{\rm em} + \mcc{L}_p$ that oscillate at the high frequency $\smash{\omega^{({\rm x})} \approx \omega^{({\rm ebw})}}$ do not contribute to the action integral $\mc{S}$ significantly on scales of interest. Thus, we can replace $\mcc{L}$ with its average:
\begin{gather}\label{eq:mccL}
\mcc{L} = \favr{\mcc{L}_{\rm em}}_{t, x} + \favr{\mcc{L}_p}_{t, x}.
\end{gather}
Here $\favr{\cdots}_t$ denotes the temporal average over the UH oscillations, $\favr{\cdots}_x$ denotes the spatial average over all oscillations, and, clearly, $\favr{\mcc{L}_{\rm em}}_{t, x} = \favr{\mcc{L}_{\rm em}}_{x}$. 

As a slow function, $\favr{\mcc{L}_p}_{t, x}$ can contain only even powers of the UH field $\smash{\vec{E}^{({\rm uh})}}$. We neglect all powers higher than the second one, assuming that waves are close to linear. (For extensions of this approach to nonlinear waves, see \Refs{my:sharm, my:itervar}.) Then,
\begin{gather}\label{eq:aux1}
\favr{\mcc{L}_p}_{t, x} = \favr{\mcc{L}_0}_x + \frac{1}{8\pi} \,\favr{\mc{U}}_{t,x}.
\end{gather}
Here, $\mcc{L}_0$ is the zeroth-order term independent of $\vec{E}^{({\rm uh})}$ and $\smash{\mc{U} = [\hat{\vec{Q}} \cdot \vec{E}^{({\rm uh})}]^2}$, where $\hat{\vec{Q}}$ is some, generally nonlocal, operator. Using the complex notation, one can also express $\favr{\mc{U}}_{t,x}$ as follows:
\begin{gather}
\favr{\mc{U}}_{t,x} = \frac{1}{2} \,\avr{
\vec{E}_{\rm c}^{({\rm uh})*} \cdot \hat{\vec{\chi}} \cdot \vec{E}_{\rm c}^{({\rm uh})}
}_{x},
\end{gather}
where $\smash{\vec{E}_{\rm c}^{({\rm uh})} \doteq \vec{E}_{\rm c}^{({\rm x})} + \vec{E}_{\rm c}^{({\rm ebw})}}$ and $\smash{\hat{\vec{\chi}} \doteq \hat{\vec{Q}}^\dag \cdot \hat{\vec{Q}}}$ is a Hermitian operator. Let us decompose $\smash{\hat{\vec{\chi}}}$ into the slow part $\smash{\hat{\vec{\chi}}_0}$ that is independent of the LH field and the remaining part $\smash{\hat{\vec{\chi}}^{({\rm int})}}$. Since $\mc{U}$ is assumed small, we adopt that $\smash{\hat{\vec{\chi}}^{({\rm int})}}$ is \textit{linear} in the LH field. This implies $\smash{\hat{\vec{\chi}}^{({\rm int})} = \Xi(\_\,, \_\,, \vec{E}^{({\rm lh})})}$ \cite{foot:ph}, where $\smash{\Xi}$ is a real rank-3 tensor that is symmetric in its first two arguments. Hence, 
\begin{gather}\notag
\hat{\vec{\chi}}^{({\rm int})} = 
\frac{1}{2}\,\big[\hat{\vec{\chi}}_{\rm c}^{({\rm int})} + \hat{\vec{\chi}}_{\rm c}^{({\rm int})}\big],
\quad
\hat{\vec{\chi}}_{\rm c}^{({\rm int})} \doteq \Xi(\_\,, \_\,, \vec{E}^{({\rm lh})}_{\rm c}).
\end{gather}
This leads to
\begin{gather}
\favr{\mc{U}}_{t,x} = \frac{1}{2} \,\text{Re}\,\big[
\vec{E}_{\rm c}^{({\rm x})*} \cdot \hat{\vec{\chi}}^{({\rm int})}_{\rm c} \cdot \vec{E}_{\rm c}^{({\rm ebw})}\big].
\end{gather}
Likewise, we can decompose $\favr{\mcc{L}_0}_x$ into terms that are of the zeroth and second order in the LH field:
\begin{gather}
\favr{\mcc{L}_0}_{x} = \mcc{L}_{p0} + \frac{1}{8\pi} \,\avr{\vec{E}^{({\rm lh})} \cdot \hat{\vec{\chi}}^{({\rm lh})} \cdot \vec{E}^{({\rm lh})}}_{t,x}.
\end{gather}
Since $\mcc{L}_{p0}$ does not depend on any of the wave variables, it does not contribute to ELEs for the wave fields (even though it contributes to ELEs for plasma particles) and thus can be omitted. Hence, \Eq{eq:mccL} becomes
\begin{gather}
\mcc{L} = \mcc{L}^{({\rm x})} + \mcc{L}^{({\rm ebw})} + \mcc{L}^{({\rm lh})} + \mcc{L}^{({\rm int})},\notag\\
\mcc{L}^{(q)} = \frac{1}{16\pi}\,\Big\{
 |\vec{E}_{\rm c}^{(q)}|^2 - |\vec{B}_{\rm c}^{(q)}|^2
 + 
 \vec{E}_{\rm c}^{(q)*} \cdot \hat{\vec{\chi}}^{(q)} \cdot \vec{E}_{\rm c}^{(q)}
 \Big\},\notag\\
\mcc{L}^{({\rm int})} = \frac{1}{16\pi}\,\text{Re}\Big\{
\vec{E}_{\rm c}^{({\rm x})*} \cdot \hat{\vec{\chi}}^{({\rm int})}_{\rm c} \cdot \vec{E}_{\rm c}^{({\rm ebw})}\Big\},\notag
\end{gather}
where we introduced
\begin{gather}
\hat{\vec{\chi}}^{(q)} = \left\{
\begin{array}{ll}
\hat{\vec{\chi}}_0,& q = \text{X}, \text{EBW},\\
\hat{\vec{\chi}}^{({\rm lh})}, & q = \text{LH}.
\end{array}
\right.
\end{gather}

Using Faraday's law and the notation defined in \Eq{eq:wk}, one gets $\smash{\vec{B}_{\rm c}^{(q)} = \hat{\omega}^{-1}\,\hat{\vec{k}} \times \vec{E}_{\rm c}^{(q)}/c}$. Also consider $\smash{\hat{\vec{\epsilon}}_0 \doteq \vec{1} + \hat{\vec{\chi}}_0}$, where $\vec{1}$ is the unit operator. Then, one can cast $\smash{\mcc{L}^{({\rm x})}}$ and $\smash{\mcc{L}^{({\rm ebw})}}$ as follows:
\begin{gather}
\mcc{L}^{(q)} = \frac{1}{16\pi}\,\vec{E}_{\rm c}^{(q)*} \cdot \hat{\vec{\mcc{D}}} \cdot \vec{E}_{\rm c}^{(q)},\label{eq:ld}
\end{gather}
where $\hat{\vec{\mcc{D}}}$ is given by \Eq{eq:D}. By treating $\smash{(\vec{E}_{\rm c}^{({\rm x})}, \vec{E}_{\rm c}^{({\rm x})*}, \vec{E}_{\rm c}^{({\rm ebw})}, \vec{E}_{\rm c}^{({\rm ebw})*})}$ as independent variables \cite{my:wkin}, one then arrives at the following ELEs:
\begin{align}
\delta \vec{E}_{\rm c}^{({\rm x})*}: & \quad \hat{\vec{\mcc{D}}} \cdot \vec{E}_{\rm c}^{({\rm x})} 
                                     = \frac{1}{2}\,\hat{\vec{\chi}}^{({\rm int})}_{\rm c} \cdot \vec{E}_{\rm c}^{({\rm ebw})},\notag\\
\delta \vec{E}_{\rm c}^{({\rm ebw})*}: & \quad \hat{\vec{\mcc{D}}} \cdot \vec{E}_{\rm c}^{({\rm ebw})} 
                                     = \frac{1}{2}\,\hat{\vec{\chi}}^{({\rm int})\dag}_{\rm c} \cdot \vec{E}_{\rm c}^{({\rm x})}.\notag
\end{align}
By comparing these with Maxwell's equations for $\smash{\vec{E}_{\rm c}^{({\rm x})}}$ and $\smash{\vec{E}_{\rm c}^{({\rm ebw})}}$, we infer that $\smash{\hat{\vec{\epsilon}}_0}$ is the plasma dielectric tensor in the operator form and $\smash{\hat{\vec{\chi}}^{({\rm int})}_{\rm c}}$ is the LHW-driven perturbation to the linear-susceptibility operator. Likewise, we infer that $\mcc{L}^{({\rm lh})}$ is the Lagrangian that determines the linear propagation of the LHW. Thus, it too can be put in the form \eq{eq:ld}. This gives
\begin{gather}\notag
\delta \vec{E}_{\rm c}^{({\rm lh})*}: \quad \hat{\vec{\mcc{D}}} \cdot \vec{E}_{\rm c}^{({\rm lh})} 
= \frac{1}{2}\,\vec{E}_{\rm c}^{({\rm ebw})*} \cdot \frac{\delta \big[\hat{\vec{\chi}}^{({\rm int})\dag}_{\rm c}\big]}{\delta \vec{E}_{\rm c}^{({\rm lh})*}} \cdot \vec{E}_{\rm c}^{({\rm x})}.
\end{gather}
In the main text, we do not use these equations \textit{per~se} but rather rederive their simplified version upon reducing the Lagrangian density $\mcc{L}$ further.



\begin{thebibliography}{10}
\bibitem{foot:ph}
The symbols $\_$ are placeholders; \ie $\smash{\vec{a} \cdot
  \hat{\vec{\chi}}^{({\rm int})} \cdot \vec{b}} = \smash{\Xi(\vec{a}, \vec{b},
  \vec{E}^{({\rm lh})})}$ for any $\vec{a}$ and $\vec{b}$.
\end{thebibliography}

\end{document}